\documentclass[aip,apl,amsmath,amssymb,reprint,superscriptaddress]{revtex4-1}
\usepackage{graphicx}
\usepackage{amsmath}
\usepackage{dcolumn}
\usepackage{bm}
\usepackage{bbold}
\usepackage{color}
\usepackage{setspace}
\usepackage{ulem}
\usepackage{nicefrac}
\usepackage{tabularx}
\usepackage{braket}

\usepackage{caption}
\usepackage{subcaption}

\begin{document} 


\title{Giant $g$-factors of Natural Impurities in Synthetic Quartz}

\author{Maxim Goryachev}
\email{maxim.goryachev@uwa.edu.au}
\affiliation{ARC Centre of Excellence for Engineered Quantum Systems, University of Western Australia, 35 Stirling Highway, Crawley WA 6009, Australia}

\author{Warrick G. Farr}
\affiliation{ARC Centre of Excellence for Engineered Quantum Systems, University of Western Australia, 35 Stirling Highway, Crawley WA 6009, Australia}

\author{Michael E. Tobar}
\affiliation{ARC Centre of Excellence for Engineered Quantum Systems, University of Western Australia, 35 Stirling Highway, Crawley WA 6009, Australia}

\date{\today}


\begin{abstract}
We report the observation of $g$-factors of natural paramagnetic impurities in a pure synthetic quartz crystal at milli-Kelvin temperatures. Measurements are made by performing spectroscopy using multiple high-$Q$ Whispering Gallery Modes sustained in the crystal. Extreme sensitivity of the method at low temperatures allows the determination of natural residual impurities introduced during the crystal growth. We observe $g$-factors that significantly differ from integer multiples of the electron $g$-factor in vacuum, and with values of up to $7.6$, which reveals much stronger coupling between impurities and the crystal lattice than in previous studies. Both substitutional and interstitial ions are proposed as candidates for the observed interactions.

\end{abstract}

\maketitle


Crystalline quartz is very important material extensively used in different areas of science and technology including optics, acoustics and device physics. In particular, unprecedented acoustic quality factors both at liquid helium and milli-Kelvin temperatures have been recently demonstrated\cite{quartzPRL,ScRep,Goryachev1}.  To further progress these areas, ultra-pure materials are required, which depend on efficient refining and identification of residual impurities. These impurities are believed to be responsible for limitations in quality factors\cite{Martin} and generation of the flicker noise as well as nonlinear effects at low temperatures\cite{quartzJAP}. In addition to applications of quartz itself, this material serves as a case study for understanding different defects in other silica-based materials.

Quartz is one of the most widely used materials due to its exceptional purity. This originates in the stable atomic configuration, which allows only few elements from the periodic table to be present in the quartz crystalline structure as an impurity\cite{mushkov}. Nevertheless, the Electron Paramagnetic Resonance (EPR) studies of this material are relatively easy due to the very narrow linewidths which increase the method sensitivity. Very narrow linewidths are explained by the fact that none of the host constitutive nuclei (most abundent isotopes) have a spin moment\cite{weil2000}. This makes quartz crystal a good candidate as a host material for Cavity Quantum Electrodynamics experiments, which interact spin ensemble impurities to microwave frequency photons\cite{ritsch, pavel1}. 

The imperfection of quartz crystals are related to substitutional and interstitial impurity ions (Al, H, Cu, Ag, Ge, P, Ti, Fe, etc) as well as vacancy centres (E$^\prime$) associated with oxygen ions missing in the crystal structure. Trivalent substitutional ions such as Al$^{3+}$, Fe$^{3+}$, Ge$^{3+}$ and Ti$^{3+}$ are typically accompanied by monovalent impurity ions, such as H$^+$, Li$^+$, Na$^+$, which are interstitially positioned in the crystal as charge compensators. There is a large number of experimental and theoretical studies dedicated to different representatives of these impurities and corresponding bonds\cite{weil1984,weil2000,mushkov}.

Quartz has been extensively studied in optical domain using various methods such as Infrared Spectroscopy, dielectric relaxation spectroscopy and thermoluminesence\cite{Lefevre}, etc. In contrast, microwave spectroscopy has been limited mostly to studies of natural quartz for geological purposes or intensionally doped or irradiated synthetic samples. Thus, more sophisticated research of synthetic quartz properties at microwave frequencies is required. In this paper, we demonstrate results of synthetic pure quartz spectroscopy in X and Ku bands ($8-22$~GHz) and at milli-Kelvin temperatures. 
 
The microwave spectroscopy has been performed using Whispering Gallery Modes (WGM) of a cylindrical shaped crystal. Due to non-negligible coupling to paramagnetic imperfections, microwave photons of such modes exhibit interaction with certain transitions of ion impurities or possibly other paramagnetic centres in a real crystal. As an effect of this coupling, WGMs exhibit considerable broadening, frequency shift or total disappearance when a transition energy of some impurity is tuned to a photon energy. The interaction is clear at low temperatures when the population of higher energy states is lower. By sweeping a DC magnetic field along the quartz cylinder symmetry axis, impurity transition energies change due to the Zeeman effect allowing observations of multiple interactions between matter and field. This spectroscopic approach has been already applied to ultra-pure sapphire crystals\cite{sapph2013} where naturally occurring ions of Fe$^{3+}$, Cr$^{3+}$ and V$^{2+}$ are identified with good agreement with theoretical predictions. 

The crystal under study is undoped crystalline quartz cylinder with the diameter of $49.9$~mm and height of $29.4$~mm. The central hole of the 4mm diameter is used to suspend the crystal with an oxygen-free copper post.    
The crystal has a seed plane normal to the cylinder axis at its centre. Since the crystal under study is undoped, only naturally occurring impurities are present. 

 The quartz cylinder is cooled down to $65$~mK with a dilution refrigeration system inside a superconducting coil magnet. The cylinder is connected to the $20$mK stage of the cryocooler via an oxygen-free copper rod. The crystal does not have a dedicated cavity, so that it is covered with a shield attached to the $100$~mK stage. 

The main limiting factor of this spectroscopic approach is the bandwidths of WGMs. The higher the quality factors of electromagnetic modes, the lower the impurity concentrations that could be detected. Although crystalline quartz exhibits significantly higher dielectric losses in the microwave region at low temperatures\cite{john,krupka} than sapphire, the achieved quality factors of WGMs are sufficient to observed a number of interactions between these modes and various impurities. The typical values of quality factors of excited WGM are of the order of $10^{5}$ with a maximum $Q$ of $2\cdot 10^6$ at $20.15$~GHz.

\begin{figure}[h!]
\centering
\includegraphics[width=3.25in]{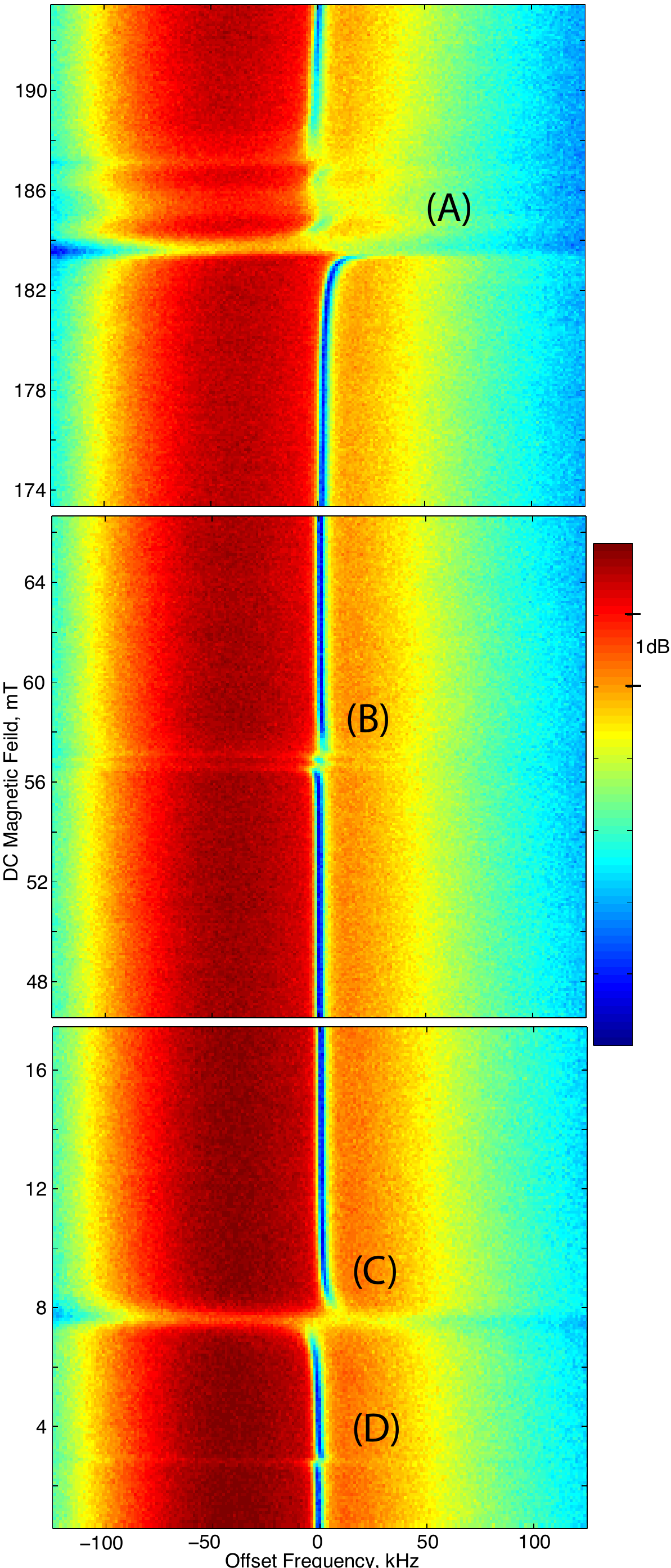}
\caption{\label{denspALL} Absolute value of the transmission through the WGM quartz resonator as a function of the excitation microwave signal and DC external magnetic field. The central frequency is $20.150079$~GHz. }
\end{figure}

\begin{figure}[h!]
\centering
\includegraphics[width=3.25in]{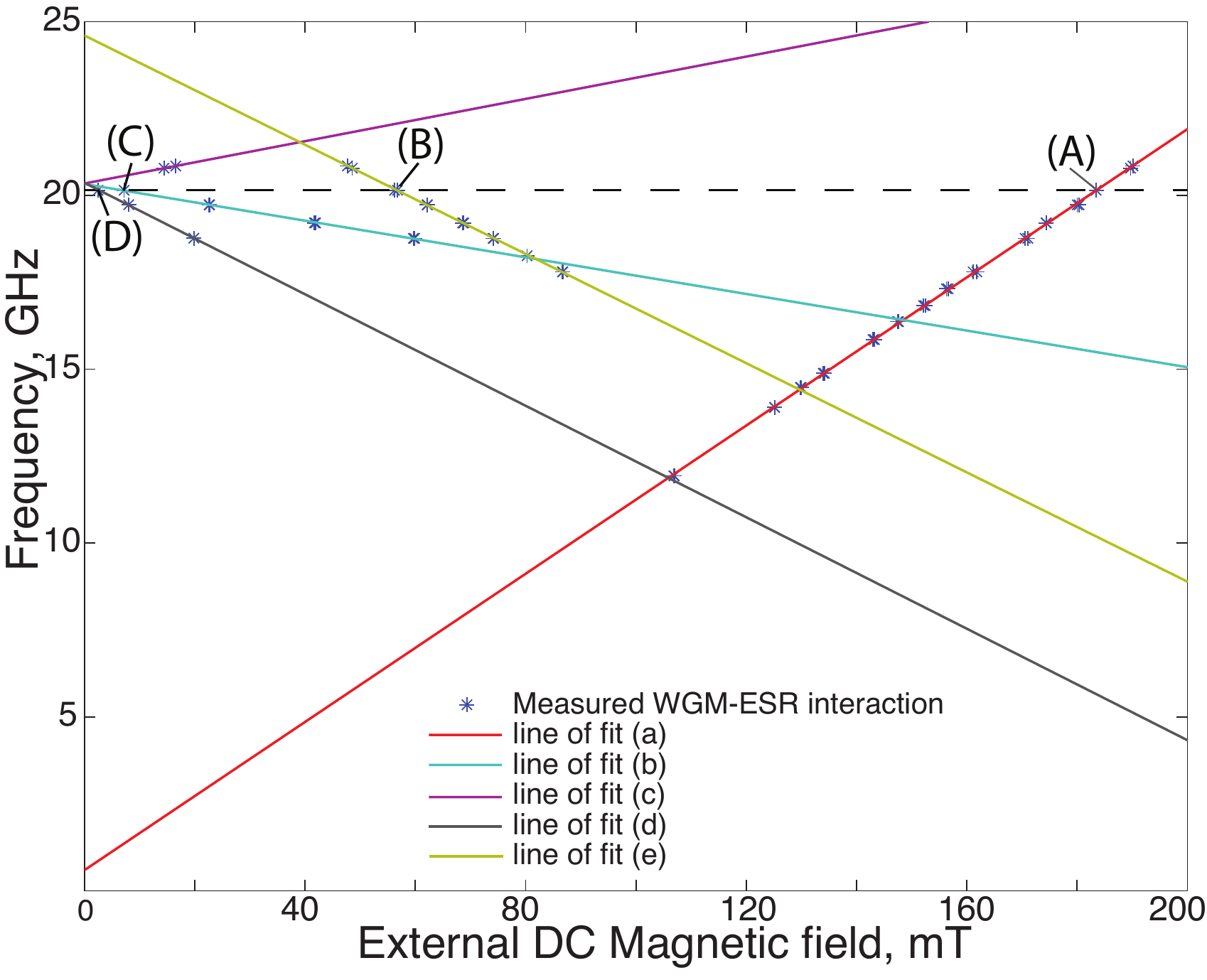}
\caption{\label{zeeman} Map of interactions between cavity WGM and paramagnetic defects. Solid lines show fitted Zeeman-like splitting lines. Parameters of the curves are shown in Table~\ref{modesT}. }
\end{figure}

The results of actual measurements are presented as density plots depicting transmission through the crystal as a function of frequency and external magnetic field. An example of interactions between a crystal WGM and paramagnetic impurities is shown in Fig.~\ref{denspALL}. The bright line at zero detuning frequency, the WGM, has field dependence only at certain values of the field corresponding to energies of some impurity transitions. A collection of WGMs gives a map of interactions in the frequency-field axes shown in Fig.~\ref{zeeman}. This figure demonstrates interactions denoted as (A)-(D) in Fig.~\ref{denspALL}. 
The dashed line denotes a WGM shown in Fig.~\ref{denspALL}. Solid lines, fitted dependencies of transition energies on magnetic field, reveal the Zeeman effect for a certain impurity. These lines are used to identify corresponding $g$-factors assuming energy levels belong to the electron. Application of the external magnetic field in the opposite direction (negative field) leads to a symmetric picture around the field-axis $B=0$. 

\begin{figure}[h!]
\centering
\includegraphics[width=3.25in]{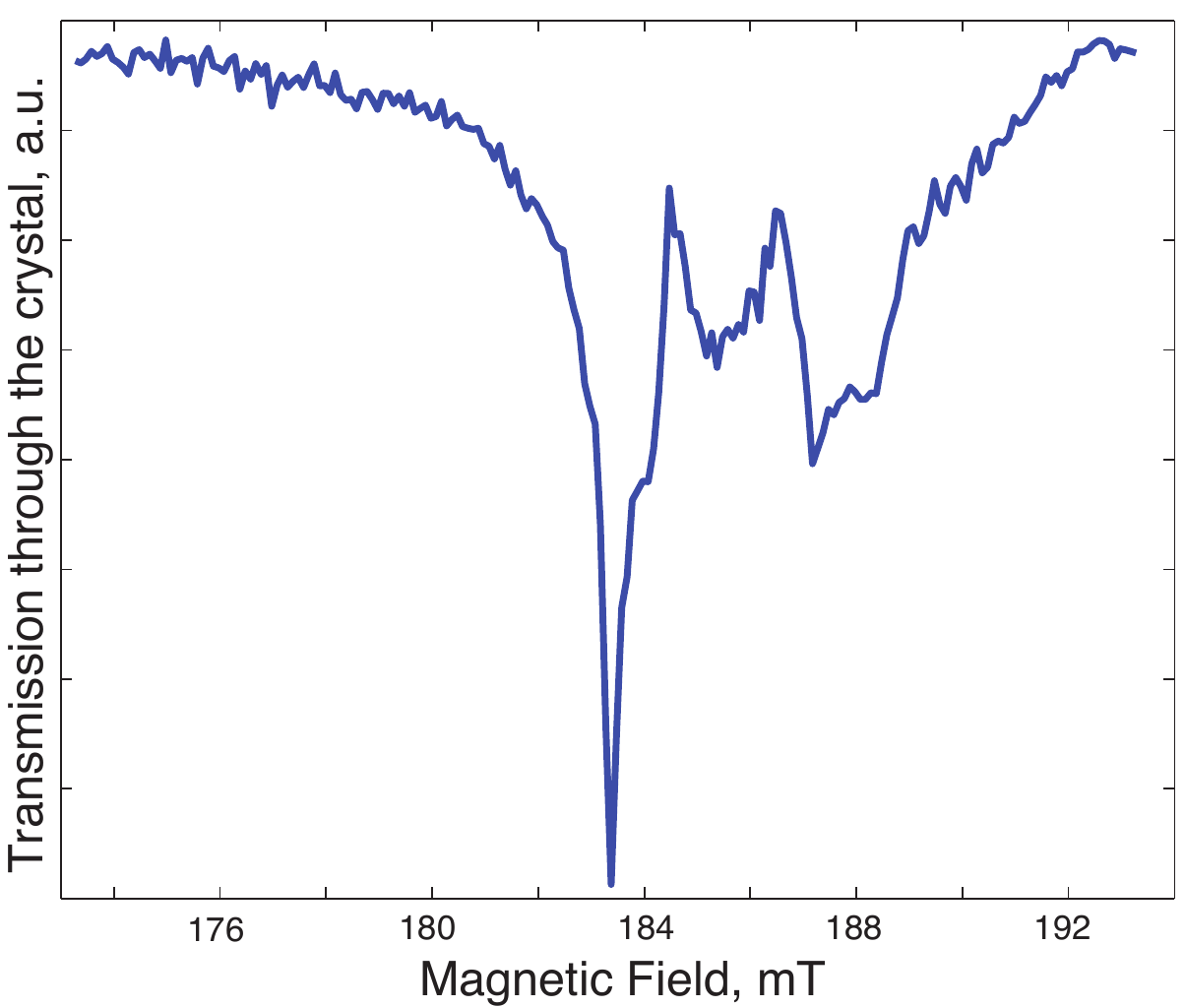}
\caption{\label{hyper} Transmission through the crystal at the WGM resonance near the interaction (A). }
\end{figure}

Interpretations of the experimental results are given in Table.~\ref{modesT}. The observed Zeeman-lines are attributed to three paramagnetic ions that are the natural impurities in the synthetic quartz crystals. Firstly, line (a) could be attributed to either lithium or nitrogen for which a hyperfine-like structure is observed (see transition (A) in Fig.~\ref{denspALL}). This structure is shown in Fig.~\ref{hyper} demonstrating three absorption dips. Such three-level system implies that the nuclear spin number of the corresponding impurity ion to be $I=1$. The only two stable isotopes with such nuclear spin are $^{14}$N and $^{6}$Li. 
$^{6}$Li ($7.6\%$ natural abundance) and $^{7}$Li ($92\%$ natural abundance) are the most suitable candidates. Another argument in favour of these interstitial ions is relatively low zero field splitting originating in week coupling to the crystal lattice.

Secondly, (b)-(d) form a transition structure that is typical for  ions with electron spin $S\ge\frac{3}{2}$ (transitions involving higher order states could be out of the observable range) and nuclear spin $I=0$, such as Fe$^{3+}$, Cr$^{3+}$ ions\cite{sapph2013}. ESR of Fe$^{3+}$ ions have been observed in synthetic brown quartz\cite{matarrese1,matarrese2} as well as studied in intentionally doped $\alpha$-quartz\cite{minge}. The most common substitutional impurity of quartz, Aluminium, also has suitable electron spin $S=\frac{5}{2}$, although nonzero nuclear spin ($I=\frac{5}{2}$) suggests the existence of hyperfine splitting that is not observed experimentally. 

Thirdly, line (e) has the almost same slope as line (d) that correspond to a two-photon transition of a previously-discussed ion with $S\ge\frac{3}{2}$ (either Fe$^{3+}$ or Cr$^{3+}$). 
Although unlike in the previous case, the ion causing interaction of line (e) has nuclear spin $I=\frac{1}{2}$ since transition (B) in Fig~\ref{denspALL} demonstrates splitting in two. Existence of this splitting does not allow us to attribute this line to $\Ket{-\frac{1}{2}}\rightarrow\Ket{-\frac{5}{2}}$ of the previous type of ion leading to attributing line (e) to a separate ion ensemble. Since line (e) corresponds to a two-photon transition, this ion possesses electron spin $S\ge\frac{3}{2}$. The only stable isotopes of such nuclear spin are $^{89}$Y, $^{107}$Ag and $^{109}$Ag that cover all 100\% of corresponding chemical elements found in nature. Traces of Ag has been determined previous in germanium-doped synthetic quartz\cite{laman}. Another explanation for line (e) is that the double line structure arises from substitution of Oxygen on two different cites with $I=0$ impurity ion.


\begin{table}[t]
\caption{Possible interpretation of fitted impurity transitions fitted in Fig.~\ref{zeeman} with fitted $g$-factors and zero field splittings (ZFS).}
\centering
\begin{tabularx}{\columnwidth}{cXXcc}
\hline
\hline
& Ions & Transition & $g$-factor & ZFS, GHz\\
\hline
\hline
(a) & $^{6}$Li$^+$/$^{7}$Li$^+$, N$^+$  & $\Ket{-\frac{1}{2}}\rightarrow\Ket{+\frac{1}{2}}$ &$7.612$ & $0.596$\\
\hline
(b) & Fe$^{3+}$, Cr$^{3+}$ &  $\Ket{-\frac{1}{2}}\rightarrow\Ket{-\frac{3}{2}}$ & $-1.883$& $20.325$\\
(c) &   $\hspace{17pt}$  & $\Ket{+\frac{1}{2}}\rightarrow\Ket{+\frac{3}{2}}$& $2.173$& $20.344$ \\
(d) &   $\hspace{17pt}$   & $\Ket{+\frac{1}{2}}\rightarrow\Ket{-\frac{3}{2}}$&$-5.609$ & $24.593$\\
\hline
(e) &   Ag$^{3+}$, Y$^{3+}$ &$\Ket{+\frac{1}{2}}\rightarrow\Ket{-\frac{3}{2}}$ &$-5.726$ & $20.355$\\
 &   O-Substitution  & & & \\
\hline
\end{tabularx}
\label{modesT}
\end{table}

Regardless of the above discussion, all Zeeman lines demonstrate anomalous values of electron $g$-factors. Indeed, all experimentally identified $g$-factors varying from $4.7$ to $8.5\%$ and are significantly different from the closest integer multiple of the electron $g$-factor in vacuum. Typically such discrepancies do not exceed $1\%$\cite{mushkov}. The values of $g$-factors observed in this work has been never reported by any of theoretical or experimental studies except for Cu$^{2+}$  ($g\approx 2.021$--$2.7$) and Ni$^+$ ($g\approx 2.088$--$2.787$) centres in specially prepared quartz\cite{solntsev}. Although neither of these two impurities fit into the observations seen here as the associated electron and nuclear spin numbers are different. The difference from the electron $g$-factor is caused by much stronger influence of the crystal lattice field on the paramagnetic impurities. 
The study also reveals a paramagnetic impurity with extremely large $g$-factor of $7.612$. These values should correspond to one photon transition since the observed interactions are extremely strong.

The present study demonstrates new experimental observations of $g$-factors in synthetic quartz spectroscopy. These results have not been predicted by any of the theoretical studies. The uniqueness of the experiment is in the extreme sensitivity of the method allowing to discover new phenomena in pure crystals at extreme low temperatures.

\begin{acknowledgments}
This work was supported by the Australian Research Council Grant No. CE110001013 and FL0992016.
\end{acknowledgments}

\section*{References}


%



\end{document}